\documentclass[apjl]{emulateapj}
\usepackage{amsfonts,amsmath,graphicx,natbib,apjfonts,lscape}

\def\cfa{1}
\def\got{2}
\def\2m1314{2M\,1314+1320} 

\begin{document}

\title{Periodic Radio Emission from the M7 Dwarf 2MASS
J13142039+1320011: Implications for the Magnetic Field Topology}

\author{ 
M.~McLean\altaffilmark{\cfa},
E.~Berger\altaffilmark{\cfa}, 
J.~Irwin\altaffilmark{\cfa}, 
J.~Forbrich\altaffilmark{\cfa}, 
and A.~Reiners\altaffilmark{\got}
}

\altaffiltext{\cfa}{Harvard-Smithsonian Center for Astrophysics, 60
Garden Street, Cambridge,MA 02138}

\altaffiltext{\got}{Universit\"at G\"ottingen, Institut f\"ur
Astrophysik, Friedrich-Hund-Platz 1, 37077 G\"ottingen, Germany}

\begin{abstract} We present multi-epoch radio and optical observations
of the M7 dwarf 2MASS J13142039+1320011.  We detect a $\sim 1$ mJy
source at 1.43, 4.86, 8.46 and 22.5 GHz, making it the most luminous
radio emission over the widest frequency range detected from an
ultracool dwarf to date.  A 10 hr VLA observation reveals that the
radio emission varies sinusoidally with a period of $3.89\pm 0.05$ hr,
and an amplitude of $\approx 30\%$ at 4.86 GHz and $\approx 20\%$ at
8.46 GHz.  The periodicity is also seen in circular polarization,
where at 4.86 GHz the polarization reverses helicity from left- to
right-handed in phase with the total intensity.  An archival detection
in the FIRST survey indicates that the radio emission has been stable
for at least a decade.  We also detect periodic photometric
variability in several optical filters with a period of $3.79$ hr, and
measure a rotation velocity of $v{\rm sin}i=45\pm 5$ km s$^{-1}$, in
good agreement with the radio and optical periods.  The period and
rotation velocity allow us to place a lower limit on the radius of the
source of $\gtrsim 0.12$ R$_{\odot}$, about $30\%$ larger than
theoretical expectations.  The properties of the radio emission can be
explained with a simple model of a magnetic dipole mis-aligned
relative to the stellar rotation axis, with the sinusoidal variations
and helicity reversal due to the rotation of the magnetic poles
relative to our line of sight.  The long-term stability of the radio
emission indicates that the magnetic field (and hence the dynamo) is
stable on a much longer timescale than the convective turn-over time
of $\sim 0.2$ yr.  If the radio emission is due to the electron
cyclotron maser process, the inferred magnetic field strength reaches
at least 8 kG.  \end{abstract}
 

\section{Introduction}
\label{sec:intro}

An unexpected result from radio observations of low mass stars and
brown dwarfs was the recent discovery of periodic radio emission tied
to the stellar rotation \citep{brr+05,had+06,hbl+07,brp+09}.
Previously thought to be incapable of generating and dissipating
strong and stable magnetic fields, observations in recent years
indicate that at least 10\% of fully convective late-M and L dwarfs
possess stable and large-scale magnetic fields
\citep{ber02,ber06,bbf+10}.  When combined with X-ray and H$\alpha$
observations \citep{mb03,rb08,bbf+10}, Zeeman broadening of FeH
molecular lines in Stokes $I$ \citep{rb07}, and spectropolarimetry in
Stokes $V$ (Zeeman Doppler Imaging: ZDI; \citealt{dfc+06}), a picture
of the magnetic topology of fully convective ultracool dwarfs is now
beginning to emerge.

Zeeman broadening observations provide a measure the average surface
magnetic field, $Bf$ (where $B$ is the magnetic field and $f$ is the
covering fraction) and have been used to study moderately rotating
objects as cool as M9 \citep{rb10}. These observations uncovered
average magnetic field strengths of up to a few kG.  The ZDI technique
has been used to study the large scale structure and topology of
magnetic fields (as it is sensitive only to the net polarization in
Stokes $V$) in objects down to spectral type M8.  At the fully
convective boundary (spectral type of $\sim {\rm M3}$) these studies
hint at a transition from non-axisymmetric toroidal fields to
predominantly poloidal, axisymmetric field topologies
\citep{dmp+08,mdp+08}, although the systematic uncertainties and
general detectability of field topologies across a wide range in
stellar parameters are not fully understood.  Similarly, there seems
to be an increase in the ratio of magnetic fluxes measured in
circularly polarized to unpolarized spectral lines at this boundary,
apparently confirming that the fully convective objects have
increasingly large-scale, low-multipole field topologies.  However, a
number of fully convective mid-M dwarfs, potentially belonging to a
younger population, still appear to possess small-scale complex fields
\citep{mdp+10}, although recent observations seem to question this
idea \citep{rei10}.

In the context of magnetic activity, the correlation between X-ray and
radio emission, which holds across several orders of magnitude in each
quantity and over a wide spectral type range \citep{gb93,bg94},
sharply breaks down around spectral type M7--M8
\citep{bbb+01,ber02,bbf+10}.  X-ray and H$\alpha$ activity also
decline precipitously around the same spectral type
\citep{mb03,rb08,bbf+10}.  Finally, the H$\alpha$ and X-ray
rotation-activity relations also break down near the M/L transition
\citep{mb03,bbg+08,rb10}.  The origin of these changes remains
unclear, in particular whether they are related to a change in the
magnetic dynamo itself, or to a change in the way magnetic energy is
released in the cool and largely neutral atmospheres.

Radio activity in ultracool dwarfs, on the other hand, appears to be
equally strong (or stronger when normalized to the bolometric
luminosity) than in early-M dwarfs \citep{ber02,ber06,bbf+10}.  Among
the active population of ultracool dwarfs, the handful of objects with
periodic radio emission exhibit diverse behavior.  In particular, the
wide variation in the temporal behavior, emission bandwidth, and
fractional polarization suggest that there are a number of different
processes at work.  Three objects (TVLM\,513-46546: \citealt{had+06};
LSR\,1835+3259: \citealt{hbl+07}; and 2M\,0746+2000: \citealt{brp+09})
have been observed to emit short-duration, highly polarized pulses
with apparently narrow bandwidth.  On the other hand, 2M\,0036+1821
exhibits periodic variability with a different structure, rising and
falling in a roughly sinusoidal manner \citep{brr+05}. In all four
objects, the observed periods, which range from two to three hours,
are in close agreement with the known rotation velocities ($v{\rm sin}
i$).

Two primary emission mechanisms have been proposed to explain the
observed radio emission: gyrosynchrotron radiation and the electron
cyclotron maser (ECM) instability \citep{ber06,tre06,had+08,brp+09}.
The short-duration periodic pulses have been attributed to a coherent
process such as ECM due their small bandwidth and high degree of
circular polarization \citep{hbl+07,brp+09}.  However, the mechanism
for the sinusoidal radio emission from 2M\,0036+1821, as well as the
quiescent emission from several ultracool dwarfs, is still under
debate, partly because of the low levels of circular polarization and
the much wider bandwidth.

Here we present multi-wavelength observations of a newly discovered
radio-emitting M7 dwarf, 2MASS J13142039+1320011 (hereafter, \2m1314),
which exhibits unique emission properties among the population of
radio active ultracool dwarfs.  It has the brightest radio emission
over the widest frequency range of any ultracool dwarf to date
(detected from 1.4 to 22.5 GHz with a nearly flat spectrum), and
presents the first clear example of sinusoidal emission (in phase at
4.96 and 8.46 GHz).  The periodicity is also detected in circular
polarization, including a helicity reversal that provides insight into
the topology of the emission regions.  Photometric observations reveal
periodic variability with the same period as in the radio band,
indicative of a large-scale magnetic spot structure.  The unique radio
properties of \2m1314, and its location at the cusp of the transition
in magnetic field properties (spectral type M7), make it an invaluable
laboratory for fully convective dynamos.

\section{Observations}
\label{sec:obs}

\subsection{Target Properties}

We identified the M7 ultracool dwarf \2m1314\ as a radio-emitting
source in 2007 during a large Very Large Array
(VLA\footnotemark\footnotetext{The VLA is operated by the National
Radio Astronomy Observatory, a facility of the National Science
Foundation operated under cooperative agreement by Associated
Universities, Inc.}) survey of nearby mid- and late-M dwarfs
\citep{mb11}.  It is a high proper motion object ($\delta{\rm
RA}\approx -244$ mas yr$^{-1}$, $\delta{\rm Dec}\approx -186$ mas
yr$^{-1}$; \citealt{ls05}), consistent with the young disk population,
and has a parallactic distance of 16.4 pc \citep{lts+09}.
Observations in the $i$ and $z$ bands with the Lucky Imaging technique
have demonstrated that it is a resolved binary with a separation of
$0.1''$ (about 1.6 AU), and a companion that is fainter by about 1 mag
\citep{lhm06}.  \2m1314\ has strong H$\alpha$ emission, with an
equivalent width of $-54$ \AA, corresponding to $L_{\rm H\alpha}/
L_{\rm bol}\approx -3.2$ \citep{lts+09}, consistent with the
rotationally-saturated level observed in early- and mid-M dwarfs.

\subsection{VLA Observations} 

We observed \2m1314\ with the VLA on several occasions and at several
frequencies (1.43, 4.86, 8.46 and 22.5 GHz).  Data were acquired as
part of programs AB1245 and AB1312 (PI: Berger).  The details of the
observations are provided in Table~\ref{tab:vla}.  All observations
were obtained in the standard continuum mode with $2\times 50$ MHz
contiguous bands.  During the 10 hr simultaneous observation we used
fourteen antennas at 4.86 GHz and twelve antennas at 8.46 GHz.  Phase
calibration was performed using J\,1309+119, while the flux density
scale was calibrated using 3C\,286 (J\,1331+305).  We reduced and
analyzed the data using the Astronomical Image Processing System
(AIPS).  We inspected the visibility data for quality, and noisy
points were removed.  In each of the observations a source coincident
with the position of \2m1314\ was detected.  To search for source
variability in the 10-hr simultaneous observation, we constructed
light curves by plotting the real part of the complex visibilities at
the position of \2m1314 as a function of time using the AIPS DFTPL
routine.

\2m1314\ is also identified in data from the Faint Images of the Radio
Sky at Twenty-Centimeters (FIRST) survey, where it was observed in
December 1999 at 1.43 GHz with a flux density of $1126\pm 123$
$\mu$Jy.  A comparison of the position in the FIRST data and in our
subsequent VLA observations in 2009 leads to a proper motion of
$\delta{\rm RA}\approx -260$ mas yr$^{-1}$ and $\delta{\rm DEC}\approx
-177$ mas yr$^{-1}$, in excellent agreement with the known proper
motion of \2m1314.

\subsection{VLBI Observations}

We observed \2m1314\ with the 10-antenna Very Long Baseline Array
(VLBA) and the Green Bank Telescope (GBT) at 8.46 GHz on 2010 April 12
UT for approximately 8 hr (program BM0327; PI: McLean).
Phase-referenced observations were carried out with a data rate of 512
Mbit s$^{-1}$ in dual polarization, using two-bit sampling.  Eight
base-band channels of 8 MHz bandwidth each were used, spread over at
total base-width of 32 MHz.  The correlator dump time was 1 s.  The
complex gain calibrator was J\,1309+1154, located at a distance of
$1.8^\circ$ from the target. There is a gap of about 50 min in the
resulting light curve of \2m1314\ due to geodetic quasar observations
intended to improve the astrometry.

\begin{figure}
\epsscale{1}
\plotone{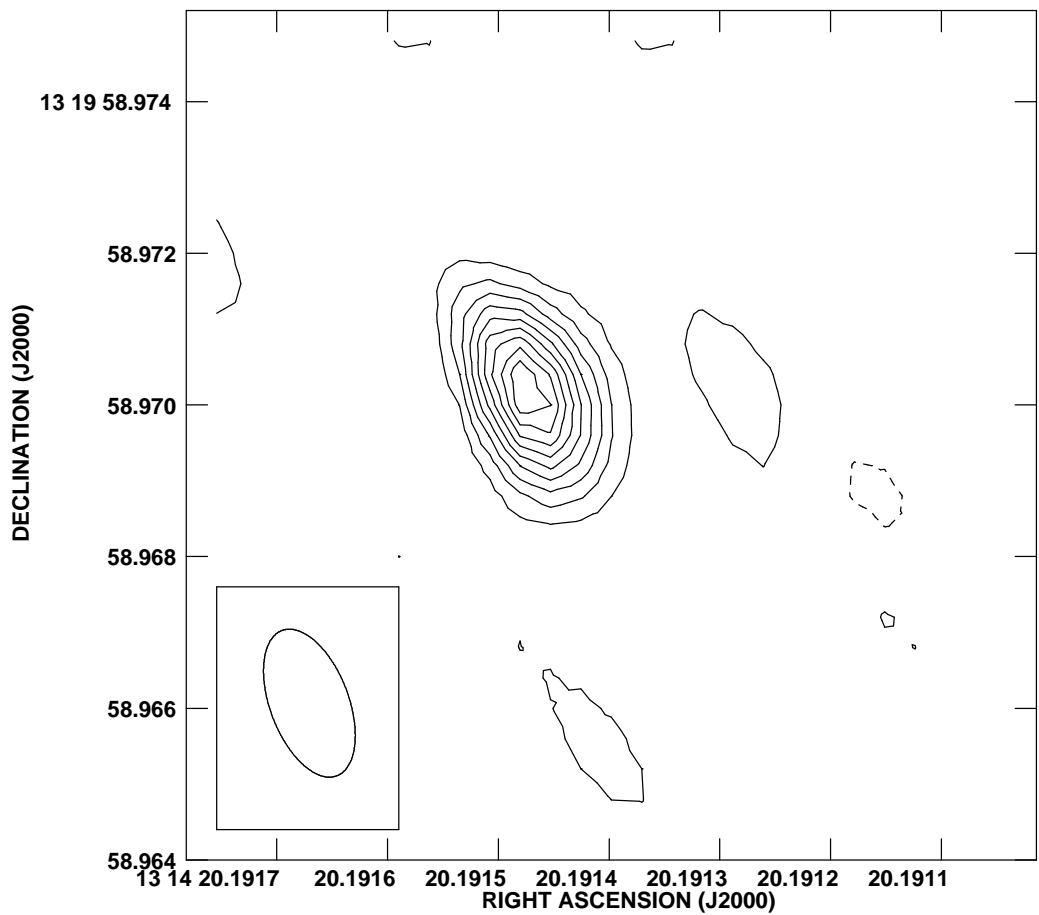}
\caption{\protect Very Long Baseline Array (VLBA) 8.46 GHz image of
\2m1314.  The source is unresolved at the VLBA scale, with a beam
size of $2\times 1$ mas (bottom left).
\label{fig:vlba}}
\end{figure}

We detected \2m1314\ with a flux density of $F_\nu=757\pm 53$ $\mu$Jy
at 8.46 GHz with $14\pm 4\%$ circular polarization.  The source is
unresolved with a beam size of $2\times 1$ mas, corresponding to an
upper limit on the source diameter of about 25 stellar radii
(Figure~\ref{fig:vlba}).  This is about an order of magnitude larger
than the expected scale of the magnetic field.  The flux density is
lower by about $27\%$ from the average of the VLA observations, most
likely due to a loss of flux caused by phase calibration errors.  We
tested this hypothesis by eliminating the longest baselines (i.e.,
reducing the angular resolution) and recovered the same flux density.
In addition, we searched for potential emission from the secondary
star in the system at a level corresponding to the missing flux
($\approx 0.3$ mJy) but we did not detect any other source above a
$4\sigma$ level of about 0.2 mJy within a $0.8''\times 0.8''$ region
centered on the detected source.

\subsection{Optical Spectroscopy}

To measure the projected rotation velocity, $v{\rm sin}i$, we obtained
a high-resolution spectrum with the Magellan Inamori Kyocera Echelle
(MIKE) spectrograph mounted on the Magellan/Clay 6.5-m telescope on
2009 May 19 UT.  The single 1000 s spectrum was reduced using a custom
reduction pipeline\footnotemark\footnotetext{\tt
http://www.ociw.edu/Code/mike} written in Python.  Wavelength
calibration was performed using ThAr arc lamps, and air-to-vacuum and
heliocentric corrections were applied.  We measured the rotation
velocity by cross-correlating the spectrum with a template spectrum of
the inactive star Gl\,876 \citep{rb08}.  The value of $v{\rm sin}i$
was determined by comparing the width of the correlation function to
cross-correlations between the template and artificially broadened
versions of the template spectrum (see for example \citealt{rb06}).
We derive $v{\rm sin}i$ from several spectral orders and find $45\pm
5$ km s$^{-1}$.  There is no evidence for a spectroscopic binary.
H$\alpha$ emission was detected with an equivalent width of 9.9 \AA,
corresponding to $L_{\rm H\alpha}/L_{\rm bol}\approx -4.0$, lower than
previous measurements \citep{lts+09}.

We also observed \2m1314\ using the Low Dispersion Survey Spectrograph
(LDSS3) on Magellan/Clay 6.5-m telescope on 2009 February 25 UT.  A
total of six 300 s integrations were obtained with the VPH-Blue grism
using a $1''$ slit.  The data were reduced using standard procedures
in IRAF, and the wavelength calibration was performed using HeNeAr arc
lamps.  We detect non-variable H$\alpha$ emission in all the
individual spectra with an average equivalent width of $\approx 14.6$
\AA, corresponding to $L_{\rm H\alpha}/L_{\rm bol}\approx -3.8$.

\subsection{Photometric Monitoring}

To study the photometric behavior of \2m1314\ we observed the source
on two consecutive nights (2010 April 6 and 7 UT) using KeplerCam on
the Fred Lawrence Whipple Observatory (FLWO) 1.2-m telescope for about
7 hr on each night.  The observations alternated between the $g$- and
$i$-band filters with individual exposures of 180 and 60 s,
respectively.  The data were reduced using standard procedures in
IRAF.  Relative photometry was performed using the Difference Imaging
Photometry Pipeline {\tt photpipe} \citep{rsb+05}.

Longer term photometric monitoring (2009 January 11 to 2010 March 18
UT) was performed with a single telescope of the MEarth array at the
Fred Lawrence Whipple Observatory.  MEarth consists of $8$ identical
0.4-m robotic telescopes, and is dedicated to a search for super-Earth
exoplanets in the habitable zones of nearby mid- to late-M dwarfs
\citep{nc08}. All observations were taken with a $715$ nm long-pass
filter, with the red end of the bandpass defined by the CCD quantum
efficiency curve.  Exposure times for data taken between 2009 January
11 and June 23 UT were $28$ s with the telescope defocused to $6$
pixels FWHM; data between 2009 December 5 and 2010 March 18 UT were
obtained with an exposure time of $11$ s with the telescope operated
in focus.  Observations were scheduled at the default $20$ min cadence
until 2009 June 12 UT when the cadence was increased to $10$ min.  On
2010 February 14 UT, we observed for the entire night at the highest
possible cadence ($\approx 30$ s, including overheads).

Basic reductions and light curve generation were performed using an
automated pipeline, based on the Monitor project pipeline
\citep{iia+07}.  A number of instrument-specific refinements have been
made, and these will be described in full in a forthcoming publication
(Berta et al. 2011, in prep).  MEarth data have two known systematic
effects remaining after the standard pipeline differential photometry
corrections, which are of particular importance for the detection of
continuous photometric modulations such as rotation, and are detailed
in \citet{ibb+11}.  These are the ``meridian offset'' and ``common
mode'' effects.  We follow the same method here to correct for these
effects, fitting for them simultaneously with the source modulations.

\section{Properties of the Radio Emission} 
\label{sec:rad}

\begin{figure}
\epsscale{0.8}
\plotone{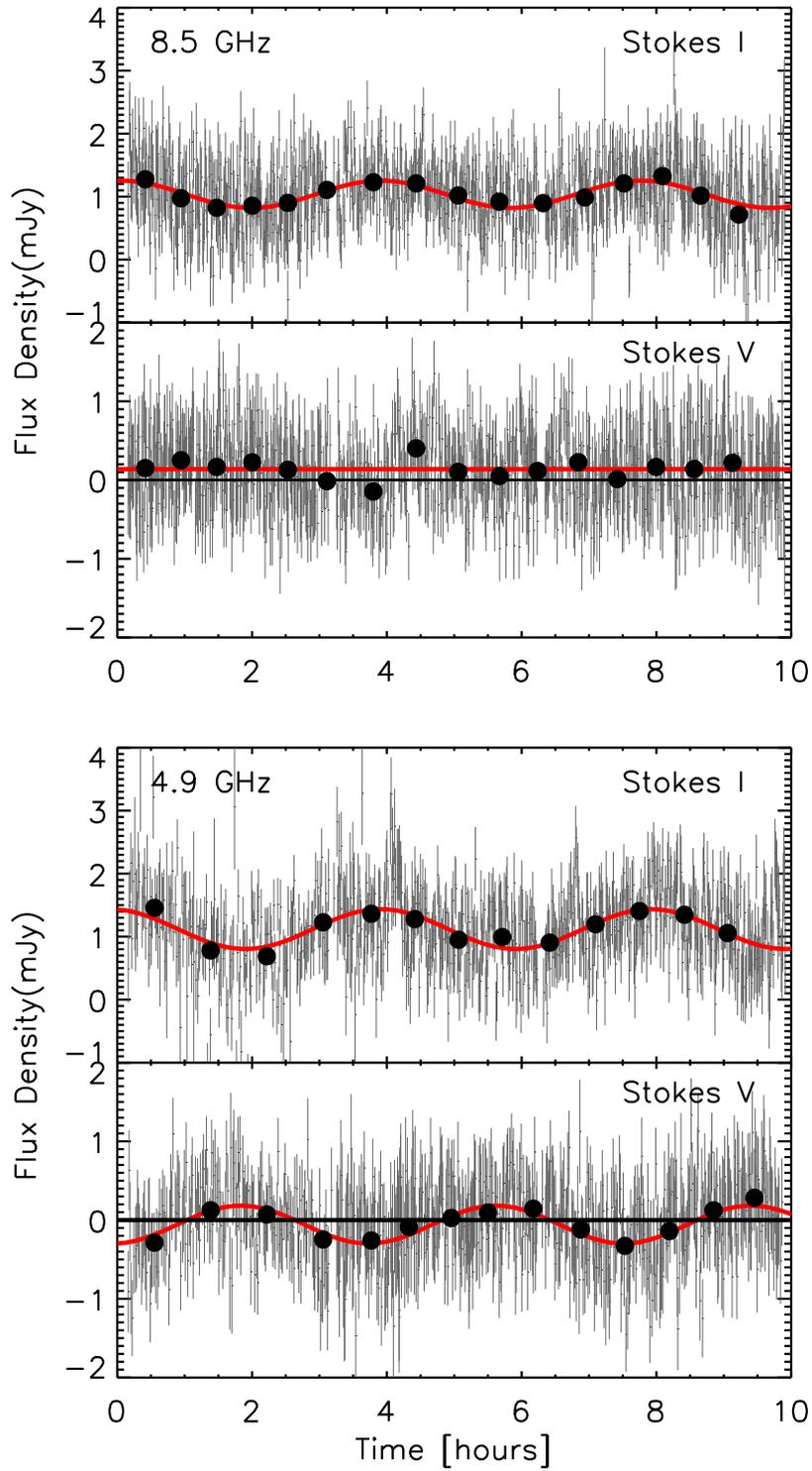}
\caption{\protect Radio light curves for the 10 hr observation of
\2m1314\ at 8.46 GHz (top) and 4.86 GHz (bottom).  The top panel at
each frequency shows the total intensity (Stokes I) while the bottom
panels display the circularly polarized flux (Stokes V).  The 30-s
averaged data are shown in grey, with 25-min averages shown as black
circles.  Sinusoidal periodicity is clearly visible at both
frequencies.  The best fit sinusoidal models are show in red,
corresponding to a $22\pm 3\%$ variation with a $3.9\pm 0.1$ hr period
at 8.46 GHz, and $32\pm 4\%$ variation with a $4.0\pm 0.1$ hr period
at 4.86 GHz.  The polarization light curve at 8.46 GHz does not show
significant periodic variation, while at 4.86 GHz it alternates
between right- and left-handed circular polarization over a best fit
period of $3.8\pm 0.4$ hr with an amplitude of $24\pm 10\%$.
\label{fig:lc}}
\end{figure}

The two initial 1-hr observations of \2m1314\ at 8.46 GHz demonstrate
a stable flux density level of about $1150$ $\mu$Jy and stable
circular polarization of about $20\%$.  Compared to the FIRST
detection in 1999 (at 1.43 GHz), the flux density level appears to be
stable for a period of about 10 yr.  The 10-hr simultaneous
observation obtained about 1.5 yr later allows us to study the
short-term temporal variability of the radio emission.  The light
curves are shown in Figure~\ref{fig:lc}.  The average flux density is
$1099\pm 18$ $\mu$Jy at 4.86 GHz and $1032\pm 16$ $\mu$Jy at 8.46 GHz,
again indicative of stable long-term radio emission, as well as a flat
spectrum.  The corresponding average luminosities are $L_\nu (4.86)=
(3.31\pm 0.05)\times 10^{14}$ and $L_\nu(8.46)=(3.52\pm 0.06)\times
10^{14}$ erg cm$^{-2}$ s$^{-1}$ Hz$^{-1}$.  Assuming that the emission
comes from the primary object, the ratio relative to the bolometric
luminosity is ${\rm log}\,(\nu L_\nu/L_{\rm bol})\approx -5.9$.  This
value is about 5 times larger than the brightest radio-emitting M
dwarfs detected to date, and roughly comparable with the L3.5 dwarf
2M\,0036+1821 \citep{brr+05}.

\begin{figure}
\epsscale{1}
\plotone{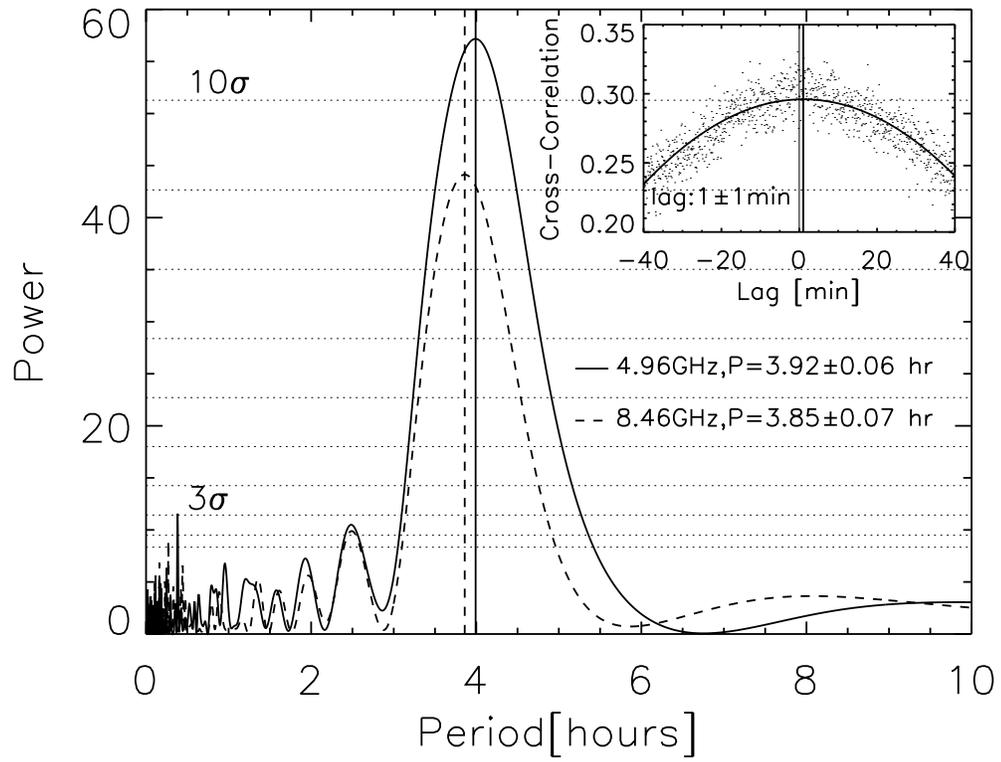}
\caption{\protect Lomb-Scargle periodograms for the total intensity
data at 4.86 GHz (solid line) and 8.46 GHz (dashed line).  A strong
peak is seen at $3.92\pm 0.06$ hr and $3.85\pm 0.08$ hr, respectively.
Significance levels of $1-10\sigma$ are plotted as horizontal lines.
Also shown is a cross-correlation of the data from both frequencies
(inset), with a peak at $1\pm 1$ min indicating no discernible lag
between the two light curves.
\label{fig:ls}}
\end{figure}

The key feature of the 10 hr observation is the clear periodic
variation in the flux density with $P\approx 4$ hr.  Unlike the
periodic radio emission detected in previous ultracool dwarfs
(TVLM\,513-46546, LSR\,1835+32, and 2M\,0746+2000;
\citealt{had+06,hbl+07,brp+09}), the variation in flux density is
smooth and sinusoidal rather than consisting of short-duration pulses.
A Lomb-Scargle periodogram of the data is shown in
Figure~\ref{fig:ls}, revealing a single significant peak at both
frequencies with $P=3.92\pm 0.06$ hr (4.86 GHz) and $3.85\pm 0.08$ hr
(8.46 GHz).  This is consistent with a single period of 3.89 hr at
both frequencies.  Cross-correlation of the two light curves shows
that there is no detectable phase shift between the two frequencies,
with a $3\sigma$ upper limit of $\approx 3$ min, or about $0.01\,P$
(Figure~\ref{fig:ls}).  We also fit the light curves with a
sine-function using a least-squares algorithm, and find $P=4.0\pm 0.1$
hr (4.86 GHz) and $3.9\pm 0.1$ hr (8.46 GHz).  The amplitude of the
variation at 4.86 GHz is $32\pm 4\%$ and at 8.46 GHz it is $22\pm
3$\%.  We do not detect obvious periodicity during the 8 hr VLBA
observation, but note that the noise level in the light curve
precludes the detection of variability at the level $20\%$ as observed
in the VLA observations.

\begin{figure}
\epsscale{1} 
\plotone{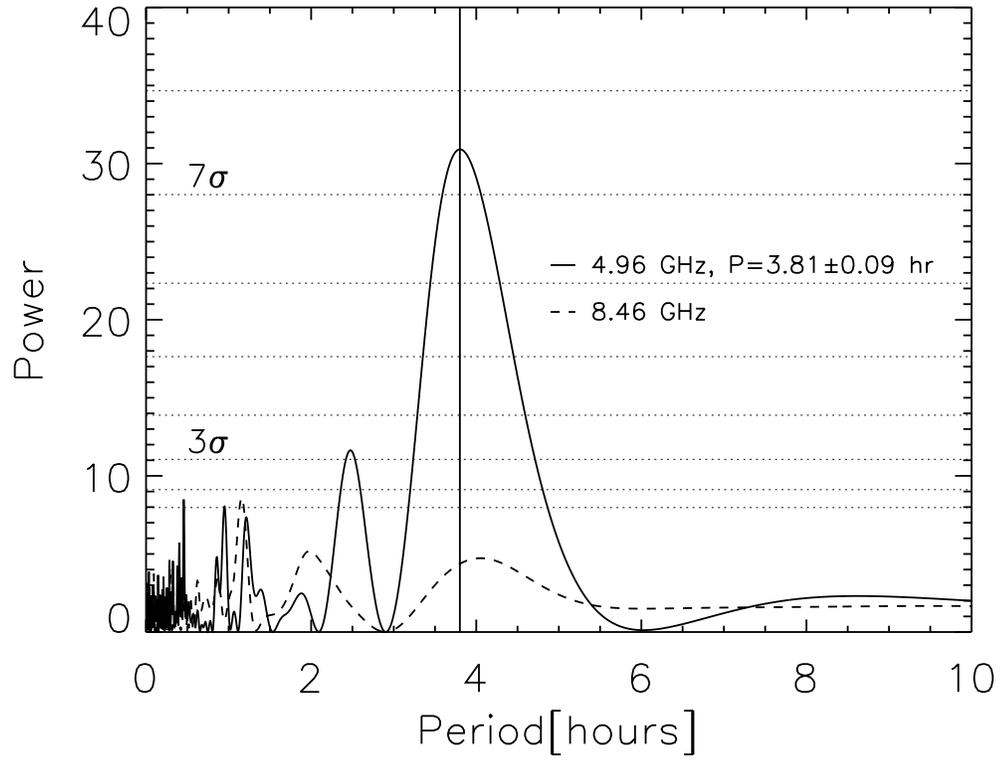}
\caption{\protect Same as Figure~\ref{fig:ls} but for the circular
polarization light curves.  A strong peak is seen at $3.81\pm 0.09$
hrs in the 4.86 GHz data, while no significant periodicity is evident
at 8.46 GHz. Significance levels of $1-8\sigma$ are plotted as
horizontal lines.
\label{fig:lspol}}
\end{figure}

The periodic variability is also detected in circular polarization.
At 4.86 GHz, the average polarization during the 10 hr observation is
close to zero but it varies at a level of $24\pm 10\%$ of the total
flux density, alternating between right- and left-handed circular
polarization over a best fit period of $3.81\pm 0.09$ hr from a
Lomb-Scargle periodogram analysis (Figure~\ref{fig:lspol}).  The
periodicity of the circular polarization is correlated with the
modulation of the total intensity, with the left-handed polarization
coinciding with the peaks in total intensity.  At 8.46 GHz the average
polarization is weaker, $\approx 10\%$, with no significant
periodicity, although the helicity of polarization may reverse during
the peaks of the total intensity (Figure~\ref{fig:lspol}).

\begin{figure}
\epsscale{1}
\plotone{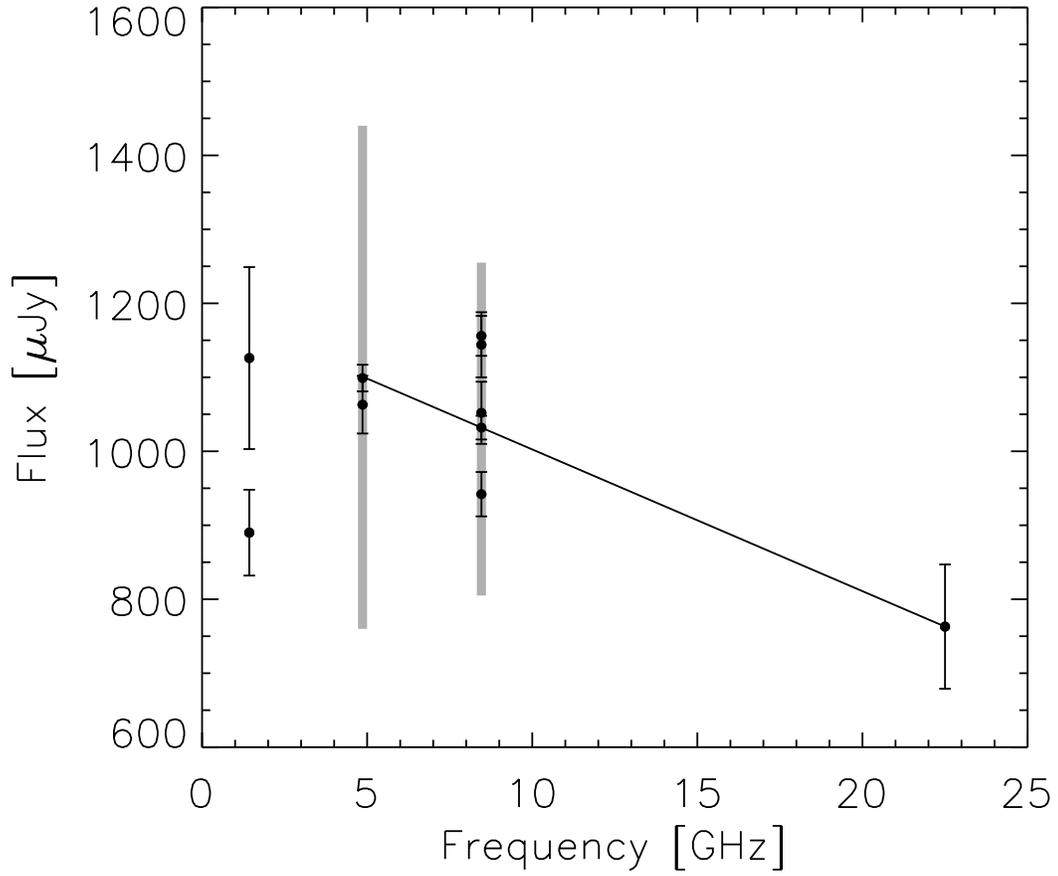}
\caption{\protect Radio spectral energy distribution of \2m1314\ from
1.4 to 22.5 GHz.  Measurements from several short observations are
shown in addition to the total flux density range detected in the 10
hour observations (vertical gray bars).  The short observations at
1.4 and 22.5 GHz provide no indication of the variability and may
not represent the mean intensity at these frequencies.  The spectrum
is surprisingly flat and appears to peak around 5 GHz with a
spectral index of about $-0.2$ at higher frequencies.
\label{fig:bb}}
\end{figure}

\begin{figure}
\epsscale{1}
\plotone{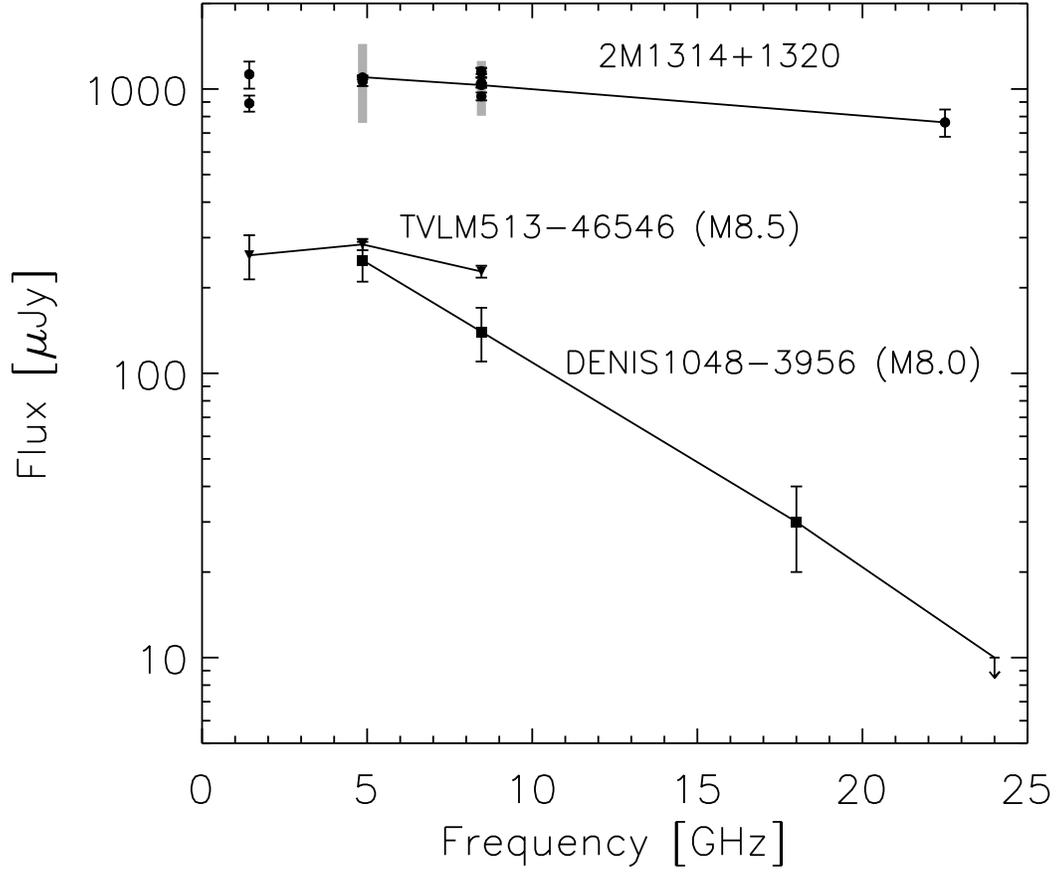}
\caption{\protect Comparison of the radio spectral energy distribution
of \2m1314\ with other ultracool dwarfs: TVLM\,513-46546 (M8.5;
\citealt{ohb+06}) and DENIS\,1048-3956 (M8; \citep{rhh+11}.  The SED
of \2m1314\ is significantly flatter than that of DENIS\,1048-3956
\label{fig:bb2}}
\end{figure}

The radio spectral energy distribution (SED) between 1.43 and 22.5 GHz
is shown in Figure~\ref{fig:bb}, including all of our VLA observations
and the FIRST survey data.  Unfortunately, only short observations
were made at 1.4 and 22.5 GHz, so we have no information about
periodicity at these frequencies.  If the flux density variations have
a similar amplitude to that observed at 4.86 and 8.46 GHz, the
measured flux may be biased low or high by about $20-30\%$ depending
on the phase of the light curve probed by these short observations.
The broad-band spectrum peaks at about 5 GHz, with a flat spectrum at
higher frequencies, $\beta\approx -0.2$ ($F_\nu\propto\nu^\beta$).
For comparison, in Figure~\ref{fig:bb2} we show the multi-frequency
data for TVLM\,513-46546 (M8.5) from 1.4 to 8.5 GHz \citep{ohb+06} and
DENIS\,1048$-$3956 (M8) from 4.9 to 24 GHz \citep{rhh+11}.  The
broad-band SED of \2m1314\ is significantly flatter than that of
DENIS\,1048$-$3956.

\begin{figure}
\epsscale{1}
\plotone{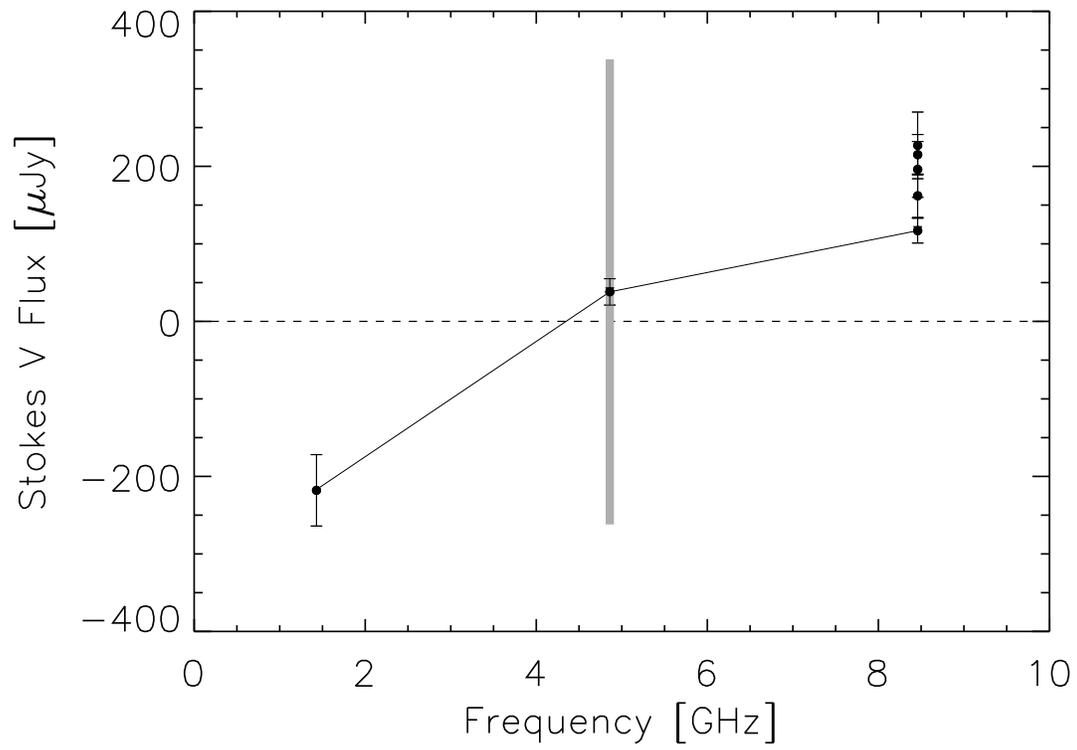}
\caption{\protect Same as Figure~\ref{fig:bb} but for the circularly
polarized flux.  It is unclear if the single 1-hr observation at 1.4
GHz represents the true mean polarization or if the object varies from
right- to left-handed circular polarization as it does at 4.86 GHz.
The polarization is consistently positive at 8.46 GHz, but is lowest
during the 10-hr observation.
\label{fig:bbpol}}
\end{figure}

The SED of circularly polarized flux is shown in
Figure~\ref{fig:bbpol}, including the full range from the 10 hr
observation.  It is unclear if the measurement from the single 1 hr
observation at 1.43 GHz represents the true mean polarization or if
the object oscillates between right- to left-handed polarization at
this frequency as it does at 4.86 GHz.  Interestingly, the
polarization is consistently positive at 8.46 GHz.  However, the 10 hr
observation has the lowest average polarization, indicating a mixture
of negative circular polarization related to the overall flux
variability.  The level of circular polarization in the VLBA detection
($+14\%$) is consistent with the VLA observations.
  
The radio emission properties can be used to extract information on
the magnetic field strength in the context of the expected emission
mechanisms.  The overall low degree of circular polarization, and the
wide frequency range over which the radio emission is detected may be
indicative of gyrosynchrotron radiation.  However, the nearly flat
spectrum corresponds to an electron power law distribution with an
energy index of $p\approx 1.3$, which is unusually low compared to
typical values of $p\sim 3-4$ \citep{gud02}.  The magnetic field
strength in the case of gyrosynchrotron emission can be estimated
using the relation $f_c\approx 2.85\times 10^3 B^{0.5}\nu_p^{-0.5}$,
indicating $B\approx 50$ G for a peak frequency of $\nu_p\approx 5$
GHz and a fraction of circular polarization of $f_c\approx 0.25$.  On
the other hand, electron cyclotron maser emission, which is produced
primarily at the fundamental electron cyclotron frequency,
$\nu_c\approx 2.8\times 10^6 B$ Hz, require the magnetic field
strength to vary from about 0.5 to 8 kG within the emission region of
\2m1314\ to explain the broad-band SED.  We note that the flat
spectrum is indicative of an even wider distribution of magnetic field
strengths.  The maximal value is much larger than the average surface
magnetic field strengths measured on similar objects using Zeeman
measurements ($1-4$ kG; \citealt{rb07}), but it is possible that the
Zeeman technique is not sensitive to such fields if their spatial
scale is relatively small.

\section{Properties of the Optical Emission}
\label{sec:opt}

\begin{figure}
\epsscale{1}
\plotone{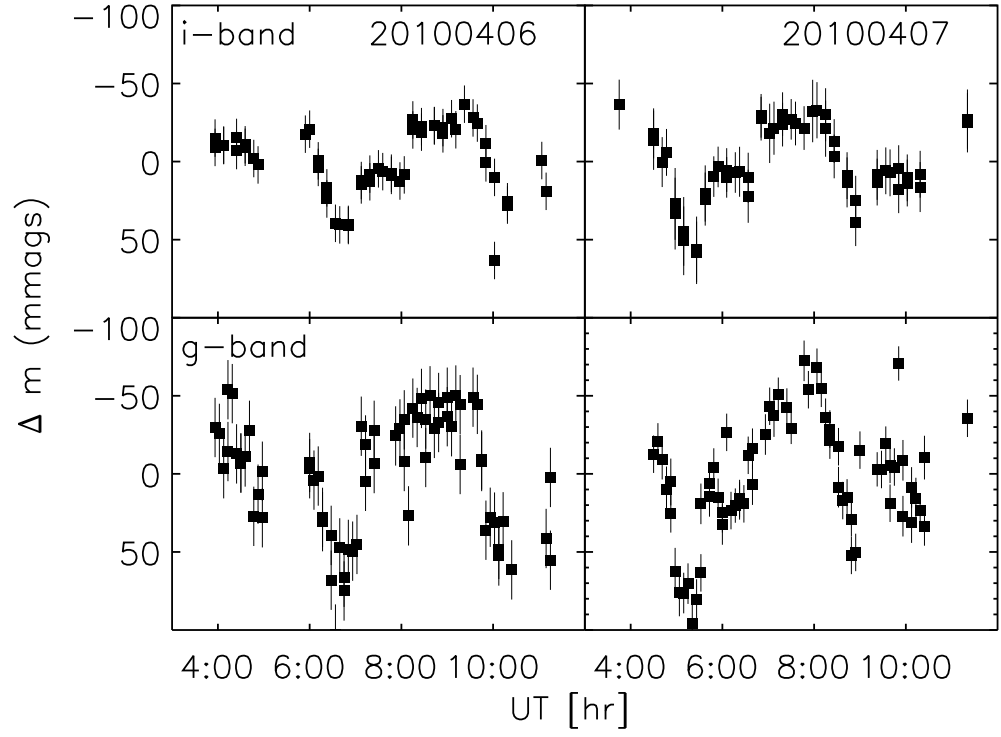}
\caption{\protect Optical $g$- and $i$-band light curves from two
nights of observations at the FLWO 1.2-m.  A period of about 4 hr is
evident.
\label{fig:flwo1}}
\end{figure}

\begin{figure}
\epsscale{1}
\plotone{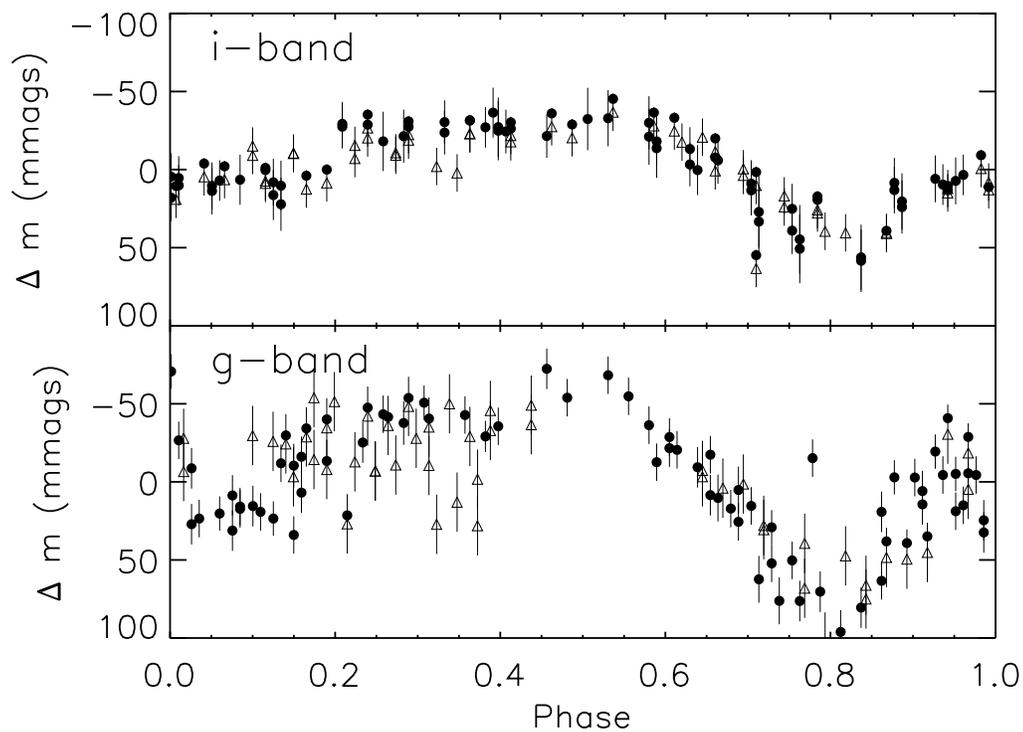}
\caption{\protect Phased light curves of the data in
Figure~\ref{fig:flwo1} folded with a period of 3.78 hr.  Data from the
two consecutive nights are plotted as triangles and circles,
respectively.  The shape of the phased light curve appears more
complex than a simple sinusoid.
\label{fig:flwo2}}
\end{figure}

\begin{figure}
\epsscale{1}
\plotone{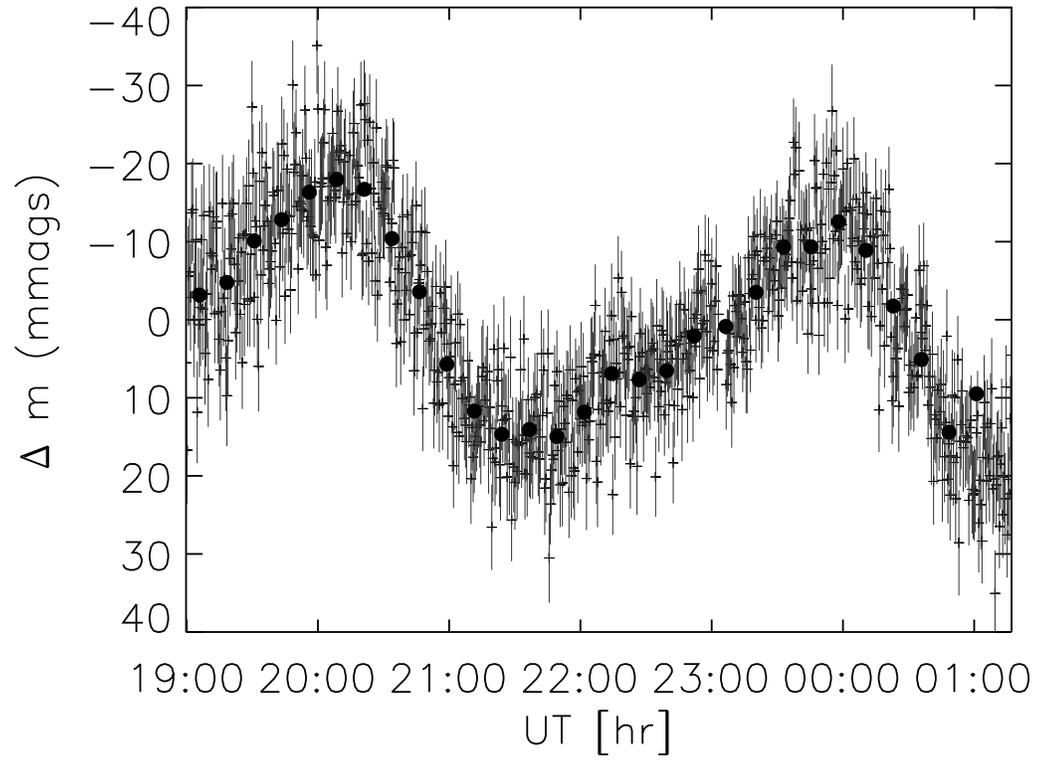}
\caption{\protect A subset of high cadence photometric observations
from the MEarth telescopes.  A period of 3.78 hr is evident with an
asymmetric light curve.
\label{fig:mearth1}}
\end{figure}

\begin{figure}
\epsscale{1}
\plotone{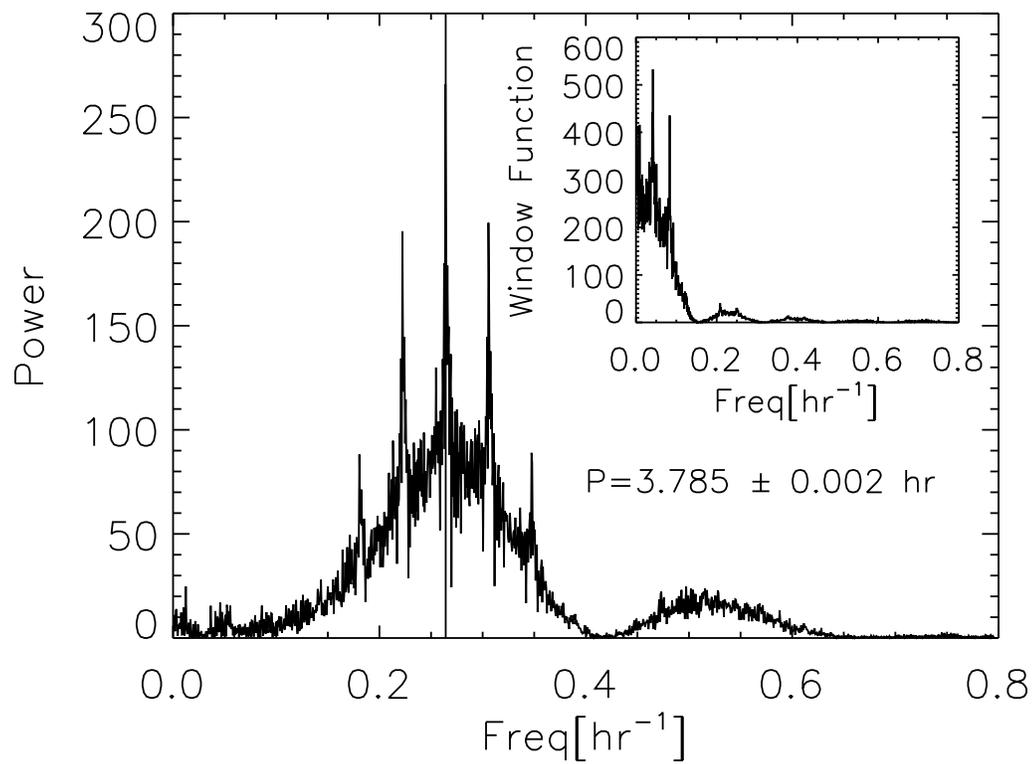}
\caption{\protect Lomb-Scargle periodogram of optical observations
from the MEarth telescopes. A clear peak is seen at a period of
$P=3.785\pm 0.002$ hr.  Aliased or harmonic peaks are also visible due
to the window function produced by the uneven sampling of the light
curve (inset).
\label{fig:mearth2}}
\end{figure}

The $g$- and $i$-band light curves from the FWLO 1.2-m observation are
shown in Figure~\ref{fig:flwo1}.  The data exhibit the same
periodicity seen in the radio, and the resulting phased light curves
(with $P\approx 3.78$ hr) are shown in Figure~\ref{fig:flwo2}.  The
variability amplitude is $\approx 80$ mmag in $g$-band and $\approx
50$ mmag in $i$-band, and both bands are in phase.  A phased light
curve from the MEarth data is shown in Figure~\ref{fig:mearth1}, with
a period of $3.785$ hr determined from a Lomb-Scargle periodogram
analysis (Figure~\ref{fig:mearth2}).  The variation amplitude in the
715 nm long-pass filter is $\approx 15$ mmag indicating an overall
trend of decreasing variability at longer wavelengths.

The phased light curve is asymmetric, with a longer rise in brightness
compared to the subsequent decline (the rise time from trough to peak
is about 2.3 hr, while the decline back to minimum takes about 1.5
hr).  The optical period is slightly shorter than the period
determined from the VLA observations, by $6\pm 3$ min.  This subtle
difference may be potentially due to differential rotation if the
radio and optical variability arise at different latitudes on \2m1314.
The MEarth and VLA observations could not be phased due to the large
gap in the observations.  The overall sinusoidal variability of
\2m1314\ points to a large covering fraction for the spot(s), while
the asymmetry is indicative of multiple or irregularly-shaped spots.

Photometric rotational modulation is common in M dwarfs and is
generally attributed to magnetic spots.  However, the cool neutral
atmospheres of ultracool are expected to also form clouds that may
produce periodic variability \citep{bai02}.  Since \2m1314\ is an M7
dwarf, its atmosphere is still fairly warm and dust clouds are less
likely.  The ratio of the variability amplitude in the $g$ and $i$
bands is about 1.6, which is consistent with a cool spot
\citep{rbm06}.  For comparison, the radio active M8.5 dwarf
TVLM\,513-46546 exhibits photometric modulation with the same period
as the radio variability, but the $g$- and $i$-band variations are
anti-correlated, indicative of clouds rather than magnetic spots
\citep{ldm+08}.  Here the $g$- and $i$-band variations are in phase,
indicating that magnetic spots are the likely explanation.  However,
it remains unclear whether the optical and radio variability are
produced by the same active regions or even related magnetic regions.

\section{Topology of the Radio Emission Region}

With a rotation period of 3.89 hr and $v{\rm sin}i=45\pm 5$ km
s$^{-1}$, the minimum radius of \2m1314\ is inferred to be $9\times
10^{9}$ cm ($0.13$ R$_\odot$).  The expected mass for an M7 dwarf
\citet{bc96} is about $0.09$ M$_\odot$, and the expected radius is
therefore about $0.1$ R$_\odot$ \citep{dsf+09}.  Thus, the inferred
radius is larger than predicted, especially if $i\ll 90^\circ$, and
this may be indicative of the trend for larger radii than model
predictions (by up to $\sim 50\%$) for magnetically active ultracool
dwarfs \citep{cgb07,lop07,jjm09}.  Given that the minimum radius is
already $40\%$ larger than the model predictions, this suggests that
that inclination is likely $\gtrsim 70^\circ$.  However, since the
binary companion is unresolved in $K$-band observations, the magnitude
of the primary is difficult to estimate.  $K$-band magnitude-mass
relations indicate the mass may actually be higher ($0.1-0.17$
M$_\odot$) and the relatively low temperature may be a result of the
inflated radius.  This may resolve some of the discrepancy between the
radius and $T_{\rm eff}$.

\begin{figure}
\epsscale{1}
\plotone{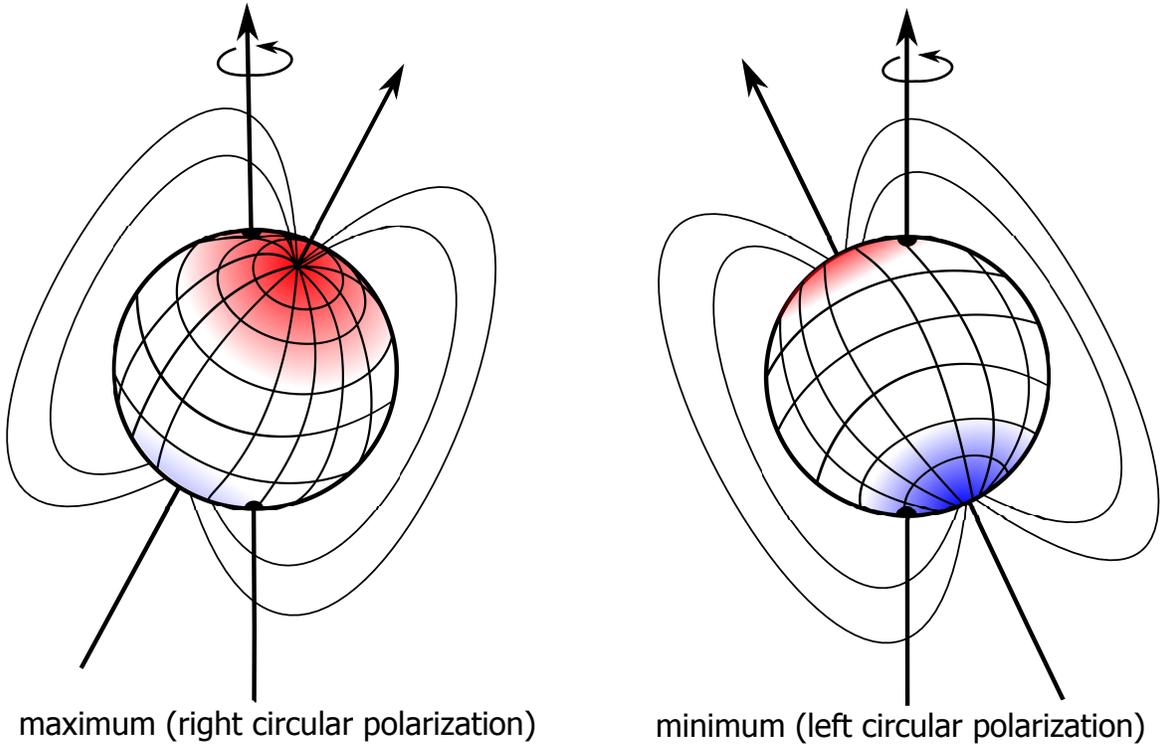}
\caption{\protect Simple geometric model of a mis-aligned dipolar
magnetic field that can explain the various properties of the observed
radio emission.  The radio emission arises from two large magnetic hot
spots with opposite polarities, located at opposite poles.  The
inclination with respect to the observer leads to a larger projected
area from one of the poles (red) and hence a peak in the total
intensity when it rotates into view.  Helicity reversals are observed
when the second pole (blue) is visible.
\label{fig:model}}
\end{figure}

The periodic radio emission, and especially the periodic helicity
reversals at 4.86 GHz, can be consistently explained with a simple
geometric model: a large-scale dipolar magnetic field mis-aligned
relative to the rotation axis and with opposite polarity at each pole;
see Figure~\ref{fig:model}.  The inclination of the magnetic axis with
respect to the observer leads to a larger projected surface area for
one of the poles, and hence an increase in the total intensity when it
rotates into view.  At 4.86 GHz we observe a helicity reversal when
the other pole is in view.  The fraction of circular polarization
remains constant, but may actually be higher than observed if the
inclination is such that both spots always remain partially visible.
If this is the case, the opposite polarities of the two spots would
cause some of the circular polarization to cancel out.  The lack of
obvious helicity reversals at 8.46 GHz could be due to changes in the
optical depth and the apparent size of the two magnetic poles as a
function of frequency, particularly if higher frequencies are emitted
near the magnetic foot-points.  This is also supported by the smaller
variability amplitude at 8.46 GHz.

The optical periodicity may be caused by cool spots spatially
connected with the radio-emitting poles, or by unrelated spots, which
are relatively common in M dwarfs.  As noted above, spatial separation
and differential rotation may explain the small difference in inferred
optical and radio periods.

\citet{mdp+10} propose two distinct magnetic topologies of mid- to
late-M dwarfs: large-scale, axisymmetric poloidal topology, and a
complex, non-axisymmetric topology.  Our geometric model indicates that
\2m1314\ has a topology more similar to the large-scale poloidal
fields rather than those with complex, small-scale fields.  The reason
for these two populations remains unclear, but \citet{mdp+10} propose
that objects may transition towards larger-scale poloidal fields with
age.  We note, however, that the objects studied in \citet{mdp+10}
have much slower rotation velocities (or correspondingly larger Rossby
numbers) than \2m1314.

\section{Conclusions} 
\label{sec:conc}

\2m1314\ provides a unique window into magnetically-induced radio
emission in ultracool dwarfs, coinciding with the spectral type at
which the radio/X-ray correlation breaks down \citep{ber02,ber06}.  It
has the most luminous stable radio emission over the widest frequency
range ($1.4-22.5$ GHz) to date, and furthermore exhibits clear
sinusoidal periodicity, with the first example of a helicity reversal
in the circularly polarized emission of an ultracool dwarf.  Taken in
conjunction, these properties point to emission from a simple dipolar
field, mis-aligned relative to the stellar rotation axis.  In the
context of electron cyclotron maser emission, the magnetic field
strength is required to be at least 8 kG.  The inferred field
structure is similar to the simple field topologies inferred from
Zeeman Doppler imaging of some fully convective dwarfs \citep{mdp+10},
while the field strength in the case of ECM emission is larger than
the typical average fields measured from Zeeman broadening of FeH
lines \citep{rb07}.

The stability of the radio emission for at least a decade indicates
that the underlying dynamo is stable well beyond the convective
turnover timescale for an M7 dwarf ($\sim 70$ d; \citealt{saa01}).
Moreover, a comparison to convective dynamo models and simulations
supports predictions of dominant large-scale, low-multipole fields
\citep{dsb06,bro08}, particularly for fast rotators \citep{bro08}, but
is at odds with models that predict small-scale, high-multipole fields
\citep{ddr93,ck06}.  

Future observations of \2m1314\ will address several key questions.
First, if the emission is due to the ECM process, we expect a cut-off
in the radio emission at a frequency that corresponds to the maximal
field strength, $\nu_{\rm max}=2.8\times 10^6\,B_{\rm max}$ Hz.
Measurements of this possible cut-off frequency will therefore provide
a crucial constraint on the magnetic field since the Zeeman techniques
are only sensitive to the average surface field or its large-scale
component.  The increased sensitivity and bandwidth of the EVLA,
particularly at frequencies of $\sim 20-40$ GHz, is well-suited for
this test.  Second, simultaneous long-term monitoring of the radio and
optical emission will test whether the variability has a similar
period and phase.  If identical, this will point to an origin in a
related magnetic field structure, while a clear difference in period
will be suggestive of differential rotation and an origin at different
latitudes.  Finally, long-term astrometric monitoring with the VLBA
will be able to uncover a companion down to a few Jupiter masses
\citep{fb09}.  Thus, further studies of \2m1314\ are bound to provide
new insights into the properties of fully convective dynamos and the
release of magnetic energy in the coolest stars and brown dwarfs.

\acknowledgements The authors would like to acknowledge assistance
from Philip Nutzman, Zachory Berta, and Peter Challis.
E.B.~acknowledges support for this work from the National Science
Foundation under Grant AST-1008361.  J.I.~gratefully acknowledges
funding for the MEarth project from the David and Lucile Packard
Fellowship for Science and Engineering (awarded to David Charbonneau),
and support from the National Science Foundation under grant number
AST-0807690.  A.R.~received research funding from the DFG as an Emmy
Noether fellow (RE 1664/4-1).  This paper includes data gathered with
the 6.5 meter Magellan Telescopes located at Las Campanas Observatory,
Chile.


\clearpage
\begin{deluxetable}{lccccc}
\tabletypesize{\footnotesize}
\tablecolumns{6} 
\tabcolsep0.15in
\tablewidth{0pt} 
\tablecaption{VLA Observations of \2m1314\
\label{tab:vla}}
\tablehead{
\colhead{UT Date}   &
\colhead{Config.}   &
\colhead{Duration}  &
\colhead{Frequency} &
\colhead{$F_\nu$}   &
\colhead{$F_{\nu,V}\,^a$} \\
\colhead{}    &
\colhead{}    &
\colhead{(hr)}  &
\colhead{(GHz)} &
\colhead{($\mu$Jy)} &
\colhead{($\mu$Jy)} 
}            
\startdata
2007 Jul 20.02 & A & 1  & 8.46 & $1156\pm 27$ & $215\pm 26$  \\
2007 Aug 10.00 & A & 1  & 8.46 & $1144\pm 44$ & $227\pm 43$  \\
2009 Mar 25.12 & B & 10 & 4.86 & $1099\pm 18$ & $38\pm 17$   \\
2009 Mar 25.12 & B & 10 & 8.46 & $1032\pm 16$ & $117\pm 16$  \\
2009 Mar 27.32 & B & 1  & 4.86 & $1063\pm 39$ & $<84$        \\
2009 Mar 27.32 & B & 1  & 8.46 & $1052\pm 42$ & $196\pm 36$  \\
2009 May 10.42 & B & 1  & 8.46 & $942 \pm 30$ & $162\pm 28$  \\
2009 May 11.37 & B & 1  & 1.43 & $890 \pm 58$ & $-218\pm 6$  \\
2009 May 11.41 & B & 1  & 22.5 & $763 \pm 84$ & $<225$ 
\enddata
\tablecomments{$^a$ Circularly polarized flux density.}
\end{deluxetable}

\clearpage

\clearpage

\clearpage

\clearpage

\clearpage

\clearpage

\clearpage

\clearpage

\clearpage

\clearpage

\clearpage

\clearpage

\end{document}